\documentclass[twocolumn]{aastex6}
\usepackage{amsmath,amsfonts,amssymb}
\usepackage{graphicx}
\usepackage{color}
\newcommand{\grs}{\object[]{GRS 1747$-$312}}

\shorttitle{Absorption edge in GRS~1747$-$312}
\shortauthors{Z.S. Li et al.}

\begin{document}

\title{Evidence for the photoionization absorption edge in a photospheric radius expansion X-ray burst from \object{GRS 1747-312} in Terzan 6}

\author{Zhaosheng Li \altaffilmark{1,2}}
%\email{lizhaosheng@xtu.edu.cn}
\author{Valery F. Suleimanov \altaffilmark{3,4,5}}
\author{Juri Poutanen \altaffilmark{5,6,7}}
\author{Tuomo Salmi \altaffilmark{6}}
\author{Maurizio Falanga \altaffilmark{2,8}}	
\author{Joonas N\"attil\"a \altaffilmark{7}}
\author{Renxin Xu \altaffilmark{9}}

\altaffiltext{1}{Department of Physics, Xiangtan University, Xiangtan, 411105, P.R. China; lizhaosheng@xtu.edu.cn}
\altaffiltext{2}{International Space Science Institute, Hallerstrasse 6, 3012 Bern, Switzerland}
\altaffiltext{3}{Institut f\"ur Astronomie und Astrophysik, Kepler Center for Astro and Particle Physics, Universit\"at T\"ubingen, Sand 1, D-72076 T\"ubingen, Germany}
\altaffiltext{4}{Kazan (Volga region) Federal University, Kremlevskaya str. 18, Kazan 420008, Russia}
\altaffiltext{5}{Space Research Institute of the Russian Academy of Sciences, Profsoyuznaya str. 84/32, 117997 Moscow, Russia}
\altaffiltext{6}{Tuorla Observatory, Department of Physics and Astronomy,  FI-20014 University of Turku, Finland}
\altaffiltext{7}{Nordita, KTH Royal Institute of Technology and Stockholm University, Roslagstullsbacken 23, SE-10691 Stockholm, Sweden}
\altaffiltext{8}{International Space Science Institute Beijing, No.1 Nanertiao, Zhongguancun, Haidian District, 100190 Beijing, China}
\altaffiltext{9}{School of Physics and State Key Laboratory of Nuclear Physics and Technology, Peking University, Beijing 100871, P.R. China}

\begin{abstract}
Thermonuclear X-ray bursts on the surface of neutron stars (NSs) can enrich the photosphere with  metals, which may imprint the photoionization edges on the burst spectra.  We report here the discovery of absorption edges in the spectra of the type I X-ray burst from the NS low-mass X-ray binary GRS 1747$-$312 in Terzan 6 during observations by the \textit{Rossi X-ray Timing Explorer}. 
We find that  the edge energy evolves from $9.45\pm0.51$ keV to $\sim6$ keV and then back to $9.44\pm0.40$ keV during the photospheric radius expansion phase and remains at  $8.06\pm0.66$ keV in the cooling tail. The photoionization absorption edges of hydrogen-like Ni, Fe, or an Fe/Ni mixture and the bound-bound transitions of metals may be responsible for the observed spectral features. The ratio of the measured absorption edge energy in the cooling tail to the laboratory value of the hydrogen-like Ni(Fe) edge energy allows us to  estimate the gravitational redshift factor $1+z=1.34\pm0.11$($1+z=1.15\pm0.09$). The evolution of the spectral parameters during the cooling tail are well described by metal-rich atmosphere models. The combined constraints on the NS mass and radius from the direct cooling method and the tidal deformability strongly suggest very high atmospheric abundance of the iron group elements and limit the distance to the source to $11\pm1$~kpc.
\end{abstract}


\section{Introduction}
\label{sec:int}

Neutron stars (NSs) in low-mass X-ray binaries (LMXBs) can accrete matter transferred from their  companions through the Roche-lobe overflow.  Depending on the mass density and temperature of the accreted gas, unstable thermonuclear burning of hydrogen, helium, or their mixture on the surface of the NSs gives rise to the phenomena known as  type I X-ray bursts \citep[e.g., see,][for reviews]{Lewin93, Strohmayer06}. Three main branches of type I X-ray bursts, i.e., normal bursts, intermediate-duration bursts, and superbursts, are distinguishable based on their bursts duration  \citep[see figure 7 in][and references therein]{Falanga08}. The light curves of these bursts showed a fast rise and approximately exponential decay with a duration of a few seconds for normal bursts, tens of minutes for intermediate-duration  bursts, and up to several hours for superbursts, depending on the ignition depth.  In particular, the superbursts are likely triggered by carbon burning in the deep NS ocean \citep{Cumming01, Zand17}.

The X-ray burst spectra can usually be well fitted by a diluted blackbody, with temperature $kT_{\rm bb}\sim1-3~ {\rm keV}$ \citep{Lewin93}. Some bursts show photospheric radius expansion  (PRE), which is believed to be a result of the luminosity approaching the Eddington value,  when the radiation pressure lifts the NS surface layers \citep{Lewin93}. During the subsequent cooling phase of PRE bursts, the photosphere is likely located at the NS surface.  This fact allows us to use PRE bursts as a tool to measure NS masses $M$  and radii $R$ \citep[e.g., see][]{Ebisuzaki87,Sztajno87,Damen90,Paradijs90,Lewin93,Ozel09,SPRW11, Poutanen14, Li15, Li17}. In particular, the NS LMXBs  in globular clusters are ideal targets to constrain the equation of state of matter at supranuclear density, because the distance to the objects can be determined independently using optical observations \citep[e.g.,][]{Kuulkers03, Guillot13, Li15, Ozel16}.

Although most of X-ray burst spectra can be well described by the Planck function, deviations from the blackbody have been observed in several cases. \citet{Zand13} found the soft and hard X-ray excesses in the spectra of an X-ray burst from SAX J1808.4$-$3658, which was interpreted as a result of enhanced persistent emission during the X-ray burst.
Many other LMXBs showed the same phenomenon \citep{Worpel13}.
Moreover, the reflection component from the photoionized accretion disk  has been observed from the superburst in 4U 1820$-$30 \citep{Ballantyne04}, 4U 1636-536 and IGR J17062-6143 \citep{Keek14, Keek17}. 

The photoionization absorption features in the range of 6--11~keV have  also been identified in 4U~0614+091, 4U~1722$-$30 and 4U~1820$-$30 \citep{Zand10}, and HETE~J1900.1$-$2455 \citep{Kajava17} due to the bound-bound or bound-free transitions of the newborn heavy elements in the X-ray burst photospheres. The observed spectral features can be associated with the hydrogen like Fe edge at 9.278~keV, or hydrogen/helium like Ni edges at 10.8/10.3~keV, respectively. 
The presence of the spectral features potentially allows to measure the gravitational redshift for the NS surface \citep{Lewin93,Ozel06} which gives additional constraints on the NS mass and radius.

GRS 1747$-$312  is a transient LMXB located in the globular cluster Terzan 6 \citep{Predehl91, Pavlinsky94, Verbunt95, Zand03, Zand03b}.  Thermonuclear X-ray bursts occurred in GRS 1747$-$312, establishing that the binary system hosts a NS. GRS 1747$-$312 is known to be an eclipsing binary in a $12.360\pm0.009~{\rm hr}$ orbit observed at an inclination of $>74.5^{\circ}$ with the eclipse duration of $\sim$ 43 min \citep{Zand00, Zand03b}. The optical counterpart has not yet been identified.

The distances to globular clusters can be measured from the relations 
  \citep{Harris96, Harris10},
\begin{equation}
(m-M)_0=V_{{\rm HB}}-M_{{\rm V}}{(\rm HB)}-3.1E(B-V),
\label{equ:dist}
\end{equation} 
where
\begin{equation}
M_{\rm V}({\rm HB})=0.16{\rm [Fe/H]+0.84}
\label{equ:hb}
\end{equation}
and $V_{\rm HB}$, $E(B-V)$ and $\rm [Fe/H]$ are referred to as the apparent magnitude of the horizontal branch stars in a globular cluster, the foreground reddening, and the metallicity, respectively. We calculate the distance to Terzan 6 by using the values and typical errors $V_{\rm HB}=22.16\pm0.20$, $E(B-V)=2.35\pm0.20$, and ${\rm [Fe/H]}=-0.62\pm0.20$ \citep{Fahlman95, Barbuy97, Kuulkers03,Valenti07}. Then, the distance uncertainty can be estimated  from Equations~(\ref{equ:dist}) and (\ref{equ:hb})  by propagating errors. % \citep{Harris10}. 
The distance modulus $(m-M)_0$ is $14.13\pm0.65$ and the distance to Terzan 6 is $6.7^{+2.3}_{-1.7}$ kpc at 1$\sigma$ conficence level. 
 
Non-Planckian spectra have also been observed in the peculiar PRE burst from GRS 1747$-$312  \citep{Zand03}. The time-resolved spectra showed a strong PRE phase occurring in the first 50 s after the  burst trigger. The spectra around the peak of the radius expansion and during the decay phase are biased from the blackbody profile since the reduced chi-square ($\chi^2_{\rm red}$) are systematically larger than 2. \citet{Zand03} added a Gaussian line to account for the possible Fe line emission, which can improve the fitting. But the Fe line has a broad feature and a low central energy of 4.9 keV. The authors claimed that this is hard to explain from Fe-K emission within reasonable shifts, especially the larger $\chi^2_{\rm red}$ appears around the peak height of the photosphere where the gravitational redshift is negligible. 

In this work, we investigate the possible presence of an absorption edge in this peculiar X-ray burst from GRS 1747$-$312 \citep{Zand03}. %The distance to  GRS 1747$-$312/Terzan 6 is given in Sect.~\ref{Sec:distance}.  
%With an additional absorption edge component, 
The analysis of the spectral properties of the burst is given in Sect.~\ref{sec:data}. We discuss the origin and variations of the absorption edge energy in Sect.~\ref{sec:edge}. The direct cooling method is applied to estimate the NS mass and radius in Sect.~\ref{sec:MR}. 
The discussion of the results is presented in Sect.~\ref{sec:Dis} and we summarize in Sect.~\ref{sec:con}.

 %In order to estimate the distance uncertainty, we adopted the values and typical errors, $V_{\rm HB}=22.16\pm0.20$, $E(B-V)=2.35\pm0.20$, ${\rm [Fe/H]}=-0.62\pm0.20$, respectively \citep{Fahlman95, Barbuy97, Kuulkers03,Valenti07}. 
%\vspace{8pt}

\section{Data Reduction}
\label{sec:data}
 The Proportional Counter Array (PCA; \citealt{Jahoda06, Shaposhnikov12}) on board the {\it Rossi X-ray Timing Explorer} ({\it RXTE}), which has an energy resolution of 17\% at 6 keV, is a powerful instrument to study X-ray bursts.
During the {\it RXTE} pointing to the source  XTE J1751$-$305 (Obs. ID. 70131-02-17-00), the PCA observed an intense type I X-ray burst, which is believed to originate from GRS 1747$-$312 located $40\arcmin$ off-axis. 
We re-analysed this  PRE burst following \citet{Zand03}. Three proportional counter units (0, 2, 3) were active during that burst. From Standard 2 PCA data, in order to guarantee roughly the  same signal-to-noise ratio for  all spectra, the data were extracted with the exposure time varying from 1 s to 4 s. Around 24 s after the burst trigger, the exposure time was set to 0.25 s to catch the rapid spectral variations. Following the method introduced by \citet{Zand03}, the correction to the collimator response was made due to off-axis pointing by {\it RXTE}. The background spectrum was extracted from 440 to 40 s before the burst trigger.   Dead-time corrections were carried out according to the instructions from the {\it RXTE} team.\footnote{\url{https://heasarc.gsfc.nasa.gov/docs/xte/recipes/pca\_deadtime.html}}
All spectra were binned to ensure each channel had at least 15 photons, and a 0.5\% systematic error was added. 

%4 s duration when the count rate was $< 1500~ {\rm cts~s^{-1}~PCU^{-1}}$, 2 s for  count rates of $> 1500~ {\rm cts~s^{-1}~PCU^{-1}}$, and 1 s for  count rates of  $\geq 3000 ~{\rm cts~s^{-1}~PCU^{-1}}$

\begin{figure}
 \epsscale{1.15}
\plotone{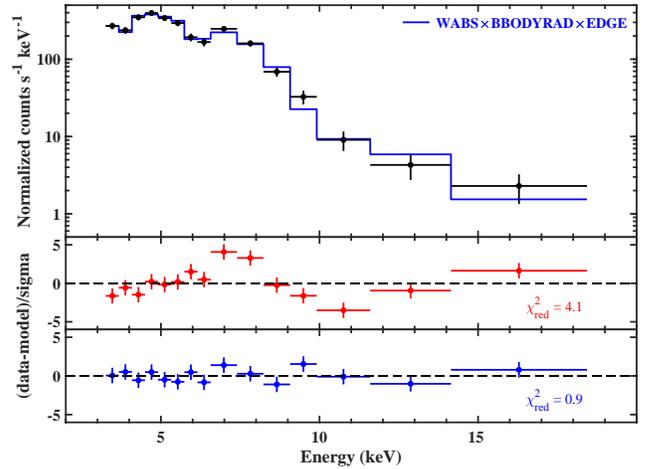}
\caption{The spectrum of the X-ray burst observed in the time interval 28--29 s after the trigger. 
In the top panel,  the {\sc wabs$\times$bbodyrad$\times$edge} model (blue stair line) is displayed, while the {\sc wabs$\times$bbodyrad} model is not shown here to avoid poor visualization. 
The residuals to the data are shown in the middle and bottom panels for the blackbody model and the edge model, respectively. 
An absorption edge improves the fit significantly.}\label{fig:spectra}
\end{figure}

The spectrum of the burst shows clear deviations from the Planckian shape. An additional absorption edge improves the fits significantly \citep[see also][]{Zand10,Kajava17}.
The edge model {\sc wabs$\times$bbodyrad$\times$edge} in {\sc xspec}  (version  12.8.2) \citep{Arnaud96} is applied to fit the spectra in the energy range 3$-$20 keV.  The blackbody component, {\sc bbodyrad}, has two parameters, the blackbody temperature, $kT_{\rm bb}$, and normalization, $K_{\rm bb}\equiv(R_{\rm bb}/D_{10 {\rm ~kpc}})^2$, to account for the emission from the photosphere, where $R_{\rm bb}$ and $D_{10 {\rm ~kpc}}$ are the blackbody radius and the distance to the source in units of 10 kpc, respectively. The multiplicative model {\sc edge} is expressed as $M(E)=\exp [-\tau (E/E_{\rm Edge})^{-3}]$ for photon energies $E>E_{\rm Edge}$ and $M(E)=1$ otherwise, where  $\tau$ and $E_{\rm Edge}$ are  the absorption depth and the threshold energy, respectively. During the fitting, we fixed the hydrogen column density at $N_{\rm H}=6\times 10^{22}~{\rm cm^{-2}}$ \citep{Zand03}. The blackbody model  {\sc wabs$\times$bbodyrad} is also used as a comparison to the edge model.  
We illustrate  in Figure~\ref{fig:spectra} how these two models fit the spectrum extracted during the PRE phase in the interval 28--29 s after the burst onset.
We see a clear improvement in $\chi^2$ and in the structure of the residuals when using the edge model. 
The bolometric flux is calculated using {\sc cflux} command in the energy range 0.001--200 keV. The $1\sigma$ errors of $kT_{\rm bb}$, $K_{\rm bb}$, $E_{\rm Edge}$, $\tau$ and flux are determined from the {\sc xspec} command {\sc error}. 
%\red{JP: can you really obtain error on flux this way?}
 
 \subsection{ Results}
 \label{sec:fit}
 
\begin{figure*}
 \centering
\includegraphics[scale=0.45]{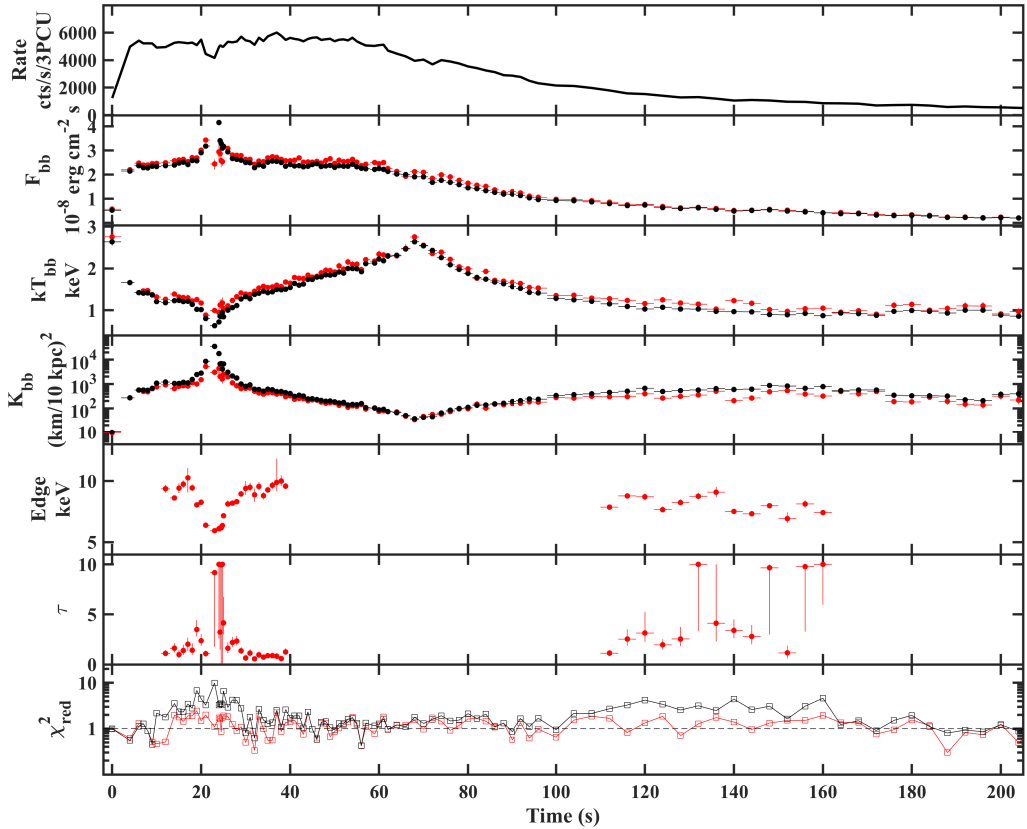}
\caption{Time-resolved spectral parameters of the PRE burst from GRS 1747$-$312, which started at the MJD 52394.04754.  From the top to the bottom panels, the dead-time-corrected light curve of PCU 0, 2, 3, the bolometric flux, the black body temperature and normalization, the absorption edge energy, the absorption optical depth, and the reduced $\chi^2$ are displayed. 
The black dots are the fitting results  using the {\sc wabs$\times$bbodyrad} model, and the red dots are from the {\sc wabs$\times$bbodyrad$\times$edge} model. 
%In the bottom panel, the dashed line shows $\chi^2_{\rm red}=1$. 
All errors correspond to $1\sigma$.}
\label{fig:burst_evol}
\end{figure*}

The results of the fitting of the time-resolved spectra are shown in Figure~\ref{fig:burst_evol}, where the red crosses and black crosses are obtained from the edge model and blackbody model respectively.  
The $\chi^2_{\rm red}$ are displayed in the bottom panel of Figure~\ref{fig:burst_evol}.  It should be noticed that the blackbody model cannot fit the data around the peak flux ($t\sim 10-40~ {\rm s}$) and the decay phase ($t\sim 110-160 ~{\rm s}$) very well, but we still plot the best-fitted values without errors.  The blackbody model fits to the data have dozens of spectra with $\chi^2_{\rm red}>2$. From the edge model, only two spectra have $\chi^2_{\rm red} > 2$ without trends in the residuals. For the multiplicative edge component, an {\it F-stat} test is appropriate to estimate the probability of chance improvement \citep{Orlandini12, DeCesar13}. We obtained the probability in the range $0.05-8\times10^{-6}$, which is a statistically significant improvement. Generally, the bolometric flux, the blackbody temperature, and the blackbody radius have small differences and almost the same trends compared with the blackbody model \citep[see also Figure 3 in][]{Zand03}. 

From the spectral fitting, the touchdown flux is $F_{\rm TD}=(2.12\pm0.07)\times10^{-8}$~erg~cm$^{-2}$~s$^{-1}$. 
%~{\rm erg~cm^{-2}~s^{-1}}$.  
Another three PRE bursts from GRS 1747$-$312 have lower touchdown fluxes. 
The two bursts observed by \textit{RXTE} \citep{Galloway08} have $(1.00\pm0.03)\times10^{-8}$ and $(1.67\pm0.04)\times10^{-8}$~erg~cm$^{-2}$\,s$^{-1}$, while the burst observed by \textit{Suzaku} \citep{Iwai14} has $(1.85\pm0.28)\times10^{-8}$~erg~cm$^{-2}$\,s$^{-1}$. Because of the strong PRE and the plateau of bolometric flux during the PRE phase, the type I X-ray burst we analyzed has been regarded as reaching its Eddington limit in previous works \citep[e.g.,][]{ Zand03, Kuulkers03, Galloway08}. The 3$\sigma$ upper limit of the persistent luminosity for our analyzed burst is $6 \times 10^{35}$~erg~s$^{-1}$ in the energy range 0.1$-$200~keV \citep[][]{Zand03}, which belongs to the quiescent state. In this case, the accretion disk is not expected to affect the NS atmosphere and the evolution of the spectra and the burst satisfies criteria from \citet{Suleimanov16}, i.e., the accretion luminosity less than  5\% of the Eddington luminosity.

 \subsection{Absorption Edge}
\label{sec:edge}

From Figure~\ref{fig:burst_evol} we see that the edge energy evolves from $9.45\pm0.51$ keV to $\sim6$ keV and then back to $9.44\pm0.40$ keV during the PRE phase and is $8.06\pm0.66$ keV in the cooling tail. A  positive correlation between the blackbody temperature and the edge energy is clearly visible.  We notice that a similar behavior of the absorption edge has been reported in 4U~0614+09. 
The  metals created by  thermonuclear explosion can produce the observed absorption edge \citep{Zand10}. 
Similarly to the case of 4U 0614+09, the bound-bound transition might be responsible for the variations \citep{Zand10}. 
Before and after the peak of the PRE phase, the edge energies of about 9.45 keV are consistent with the photoionization edge of the hydrogen-like Ni at 10.8 keV with  redshift $z_{\rm PRE}= 1.14\pm0.05$. 
During the cooling tail, the edge energy $E_{\rm Edge}=8.06\pm0.66$ keV corresponds to the redshift after the touchdown of $1+z_{\rm TD}=1.34\pm0.11$ for the hydrogen-like Ni edge, if the photosphere was located at the NS surface. 
These estimates give us a ratio $(1+z_{\rm TD})/(1+z_{\rm PRE})=1.18\pm0.11$.

Another way to estimate this ratio from the measured fluxes in the PRE phase, $F_{\rm PRE}$, and at the touchdown, $F_{\rm TD}$, which satisfy the relation $F_{\rm PRE}(1+z_{\rm PRE})=F_{\rm TD}(1+z_{\rm TD})$, if the flux in both stages, is to use the local Eddington flux.
Taking the mean PRE flux $F_{\rm PRE}=(2.69\pm0.26)\times10^{-8}$~erg~cm$^{-2}$~s$^{-1}$ we get  $(1+z_{\rm TD})/(1+z_{\rm PRE})=1.26\pm0.13$, which is consistent within 1$\sigma$ with the value obtained above assuming the hydrogen-like Ni/Fe origin of the edge.
We therefore suggest that the absorption edges in the burst spectra come from the heavy elements, i.e. hydrogen-like Ni/Fe. 
Absorption edges from a mixture of  Fe and Ni are also possible.

\section{The Mass-Radius Constraints}
\label{sec:MR}

 %The chemical composition of $Z=40Z_{\odot}$, i.e.,  $X=0.352,~Y=0.118$, $Z=40Z_\odot$ ($Z_\odot=0.0134$),  and pure iron are adopted individually in this work.  The set of hot NS atmospheres are provided for $\log g=14.0$, $\log g=14.3$, and $\log g=14.6$ \citep{Nattila15}.
We took the advantage of the direct cooling method to determine the NS mass and radius of GRS 1747$-$312  as proposed by \citet{Suleimanov17}.  
The procedure is described briefly here. We use two NS atmosphere models from \citet{Nattila15}, with the atmosphere composition of  $X=0.352$, $Y=0.118$, $Z=40Z_\odot$ ($Z_\odot=0.0134$) for the abundances of hydrogen, helium, and metals, respectively, and the pure iron model. 
%JP: remove because it makes no sense to discuss this here \textbf{ We note that the pure iron model  provides similar NS mass and radius as the pure nickel model. So we adopted the pure iron model to represent the pure nickel  or mixed nickel/iron atmosphere.} 
In these models the NS surface gravity, $\log g$, varies from 13.7 to 14.9 with a the step of 0.1 for the pure iron and with the step of 0.15 for the $Z=40Z_\odot$ models. 
The models at intermediate $\log g$ were obtained by linear interpolation.

The direct cooling method suggests minimizing the function  
\begin{equation}\label{equ:chi}
\chi^2  =   \sum_{i=1}^{N_{\rm obs}}\left[ \frac {(w \Omega - K_i)^2} {(\Delta K_i)^2}+ \frac{( w f_c^4\ell F_{\rm Edd}- F_{i})^2}{\Delta F^2_{i}}\right] .
\end{equation}
Here $w$, $f_c$, and $\ell$, are the model spectral dilution factor, the color correction factor, and the flux (or luminosity) in units of the Eddington limit, respectively; $\Omega=(R(1+z)/D)^2$ is the angular dilution factor and  $F_{\rm Edd}=GMc/(1+z)\kappa_{\rm T}D^2$ is the observed Eddington flux, where $\kappa_{\rm T}=0.2(1+X)$ is the Thomson electron scattering opacity and $X$ is the hydrogen mass fraction.  
The $K_i$ and $F_i$ are the $i$th observed blackbody normalization and flux during the cooling tail,  where $\Delta K_i$ and $\Delta{F_i}$ are the corresponding errors.  
For each observed point $(K_i,F_i)$ we take the minimum value of the normalized distance estimated in the square brackets of Equation (\ref{equ:chi}) to the dense set of points along the model curve $(w\Omega,w f_c^4\ell F_{\rm Edd})$. 
 
\begin{figure}
\epsscale{1.15}
\plotone{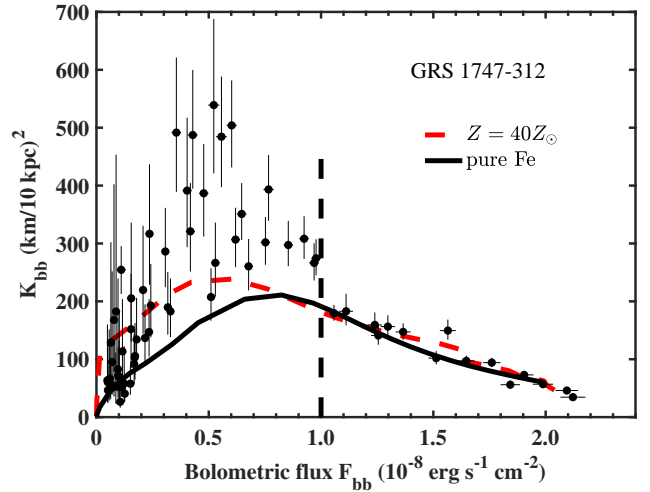}
\caption{The blackbody normalization dependence on flux for the PRE burst. The data are shown with black crosses. The solid black curve and red-dashed curve show the best-fit models for the pure iron atmosphere and $Z=40Z_{\odot}$ from \citet{Nattila15}, respectively. 
Only the data on the right hand side of the vertical dashed line are fitted. }\label{fig:xte_relations}
\end{figure}

We consider a grid of masses $M$ from 1 to 3$M_{\odot}$ with the step of $0.01M_{\odot}$ and the grid of radii $R$ from 9 to 18 km with the step of 0.01 km. 
The pairs $(M,R)$ are ignored if the causality condition $R>2.9GM/c^2$ is not satisfied \citep{Lattimer07}. 
For a given pair $(M,R)$ the surface gravity $\log g$ and the gravitational redshift $z$ are computed. 
Then the spectral dilution factor $w$ and the color correction factor $f_c$ are then determined from the theoretical models \citep{Nattila15} for the given atmosphere composition.  
We interpolate the sets of models to find the theoretical curve $w-wf_c^4\ell$ for the corresponding $\log g$. 
Then we fit the data  in the time interval $68\leq t \leq 96$~s (see Figure~\ref{fig:burst_evol}),  i.e. after the touchdown down to the flux level of $10^{-8}$~erg~cm$^{-2}$~s$^{-1}$. 
We do not use the data at lower fluxes, because  the model clearly deviates from the data (see Figure~\ref{fig:xte_relations}).
 % to avoid the possible of \textbf{extended NS atmosphere (see Sect.~\ref{Sec:extend})} at the end of the burst. % \citep[see, e.g.,][]{Nattila16, Suleimanov17b}
We also note that the selected spectra corresponding to rather high flux and temperature are well fitted with the model without the edge (see Figure~\ref{fig:burst_evol}). 
This allows us to use atmosphere model spectra that have been fitted with the blackbody only to obtain $w$ and $f_{\rm c}$.
The best-fit theoretical model computed for $\log g=14.3$ together with the observed $K_{\rm bb}-F_{\rm bb}$ relation are shown in Figure~\ref{fig:xte_relations}. 

%\red{JP: please, check the time interval}

\begin{figure}
\epsscale{1.1}
\plotone{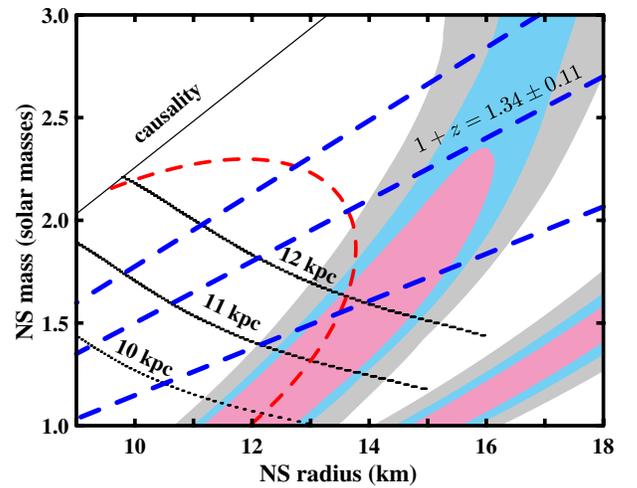}
\caption{
%Same as Figure~\ref{fig:MR1} but for the pure iron atmosphere model. 
Mass--radius constraints for the NS in GRS 1747$-$312 using the pure iron (upper contours) and $Z=40Z_\odot$ (lower right contours) atmosphere models. 
The grey, blue, and pink contours show 68\%, 90\%, and 99\% confidence regions, respectively.  
The best fit gives for 12 dof $\chi^2=14.66$ and 13.78, respectively, for the two models. 
The blue dashed lines correspond to the redshift factor $1+z=1.34\pm0.11$ measured from the absorption edges.  
The red dashed curve  shows the contour of the constant Eddington temperature $T_{\rm Edd, \infty}$ given by Equation (\ref{equ:Tinf}) for the best-fit iron model. 
The black dotted curves correspond to the constant distance of 10, 11, and 12 kpc for the iron model.
%\red{JP removed figure for $40Z_\odot$ model - too much attention to the results that make no sense; please, add contours for $40Z_\odot$ to this figure}
}
\label{fig:MR2}
\end{figure}

We note that for a given $(M,R)$ and chemical composition the only fitting parameter in Equation~(\ref{equ:chi}) is the distance to the source, $D$. 
%by minimizing $\chi^2$ in Equation~\ref{equ:chi}. 
%After we fit the theoretical curve $w-wf_c^4\ell$ to the observed $F-K_{\rm bb}$ relation
The minimum $\chi^2$ for each pair of $(M,R)$ and the global minimum together with the best-fit $D$ are obtained with the regression method. 
We find the best-fit $\chi^2= 13.78$ for $Z=40Z_\odot$ and  $\chi^2= 14.66$ for pure iron (both for 12 dof). 
In the $M-R$ plane, $\Delta\chi^2$ can be transferred to  find the confidence regions, which are shown in Figure~\ref{fig:MR2}. 
It is clear that NS radii can lie within a reasonable range of $10<R<14$~km \citep[see e.g.][]{Steiner13,Nattila17} only for models with very high metal abundance, i.e. pure iron in our case.
The obtained fits also limit the possible distance to the source (see the dotted curves in Figure~\ref{fig:MR2}).

For illustrative purposes, we can draw a curve on the $(M,R)$ plane corresponding to the observed Eddington temperature obtained from the best-fit parameters $F_{\rm Edd}=1.90\times 10^{-8}$~erg~cm$^{-2}$~s$^{-1}$ and $\Omega=199$~(km/10~kpc)$^2$ for the pure iron model shown in Fig.~\ref{fig:xte_relations}.  
From the definition of the Eddington temperature (which is independent of the distance) we get \citep{SPRW11}:
\begin{equation}\label{equ:Tinf}
T_{\rm Edd, \infty}=9.81 (F_{\rm Edd, -7}/\Omega[({\rm km/10 kpc})^2])^{1/4}~{\rm keV} 
\end{equation}
where $F_{\rm Edd, -7}=F_{\rm Edd}/10^{-7}$~erg~cm$^{-2}$~s$^{-1}$. 
The contour of constant Eddington temperature $T_{\rm Edd, \infty}=1.72$~keV is shown in Figure~\ref{fig:MR2} by the red dashed curve for the pure iron model. 
It is in a good agreement with the $M-R$ confidence regions (we do not expect that they fully overlap, because the value of $T_{\rm Edd, \infty}$ is computed for $\log g =14.3$ only and in reality varies across the $(M,R)$ plane). 
Moreover, the region of constant gravitational redshift $1+z=1.34\pm0.11$  overlaps with the 68\%, 90\%, and 99\% $M-R$ confidence regions for the pure iron model, which means a good consistency. 
However, for the $Z=40Z_{\odot}$ model, the redshift constraints are inconsistent with the $M-R$ confidence region (see the lower contours in Figure~\ref{fig:MR2}) for reasonable NS radii.

\section{Discussions}\label{sec:Dis}
 In this work, we analyze the peculiar type I X-ray burst from GRS 1747--312, which showed strong discrete feature in the burst spectra.  \citet{Zand03} fitted the feature by a broad emission line and claimed that two radius expansion phases, which correspondingly lasted 70 s for the slow one and  8 s  for the fast one, occurred in GRS 1747--312.  However, the broad emission line and the fast radius expansion phase are both difficult to explain. We found that the burst spectra are fitted well using the blackbody component with an additional photoionization edge from heavy elements. Then the presence of the fast radius expansion phase is not needed (see Fig.~\ref{fig:burst_evol}).

During the cooling tail, the edge energies showed fluctuations, which cannot be solely from the limited energy resolution of {\it RXTE/PCA}. We note that the edge energies for the H-like Fe and H-like Ni are 9.278 keV and 10.8 keV, respectively. So we may observe the gravitational redshifted edge energies between $9.278/(1+z)-10.8/(1+z)$, i.e., 7.4 -- 8.6 keV for $1+z=1.26$, for a mixture of Fe/Ni. These values are close to the fluctuations, i.e., between 7.0 $\pm$ 0.4 keV and 9.0 $\pm$ 0.4 keV. In the PRE phase, the edge energies are systematically larger than those during the cooling tail, which may indicate changes in edge energy from the gravitational redshift variations.

\subsection{Reason for the vanishing of the photoionization edge around the touchdown}

Close to the touchdown, i.e., $t\sim40-110$~s after the burst onset, the blackbody model can fit the spectra very well, which means the photoionization edge component is unnecessary. A similar  vanishing of the photoionization edge was also observed in 4U 0614+091, 4U 1722$-$30, 4U1820$-$30 and HETE J1900.1$-$2455 \citep{Zand10,Kajava17}. The absence of edges is easy to understand because at high temperatures the metals (Fe or Ni) in the NS atmosphere are fully ionized.
On another hand,  the photoionization edge appeared again $t\sim110-160$~s after the burst onset. At this time, the temperature was much lower ($\sim 1~ {\rm keV}$, see Figure~\ref{fig:burst_evol}) and the iron in the NS atmosphere was only partially ionized.  

\subsection{Possibility of an expansion phase in the cooling tail}
\label{Sec:extend}

At  fluxes smaller than $10^{-8}~{\rm erg~cm^{-2}~s^{-1}}$, the blackbody normalization, $K_{\rm bb}$, is about two times larger than the model predicts. If we interpret this as an evidence for an extended atmosphere, this would have two consequences. 
First, the radius of an atmosphere larger than $R_{\rm NS}$ would produce a larger $K_{\rm bb}$. 
If we accept the radius of the NS to be 13~km (as measured for $M_{\rm NS}=1.5M_\odot$ from the spectral evolution at higher fluxes) and  $1+z=1.23$, the atmosphere radius at low fluxes is about 18~km, corresponding to $1+z=1.15$, if the color correction factor variation is ignored. 
Second, the theoretical relation $K_{\rm bb}-F_{\rm bb}$ is derived from the plane-parallel assumption, which is not satisfied for an extended atmosphere \citep{SPW11, Nattila15}. 
A second photospheric expansion phase may arise due to an increase of the radiation pressure  when the heavy elements are not fully ionized, as we observe at the absorption edge. Because in our atmosphere model we do not take into account spectral lines and  photoionization from the excited levels, the additional sources of  opacity can, potentially, lead to atmospheric expansion.  In this case, the true gravitational redshift should be larger than the measured value. 

\subsection{Effect of ignorance of NS spin}
In this work, the effects of NS spin were ignored. Actually, the rotation of the NS can smear the absorption edge, which can be fitted by the {\sc smedge} model in {\sc xspec} \citep{Salvo09} or by the  {\sc dpsmedge} model \citep{Iwai17}. The {\sc smedge} model, with the photoionization cross-section index  fixed at $-3$, was also applied to account for the absorption. We found that $E_{\rm edge}$ is well consistent with the pure edge model, while the optical depth, $\tau$,  is slightly larger. Statistically, the {\sc smedge} cannot provide better reduced $\chi^2$ than the {\sc edge} model. Moreover, the estimation of gravitational redshift only depends on the measured energy, so ignorance of it would not affect our conclusions.

\subsection{$M-R$ of the NS in GRS 1747--312}
The spectral data for the X-ray burst cooling tail from GRS 1747$-$312 are not of sufficient quality to place tight constraints on the NS mass and radius. 
Therefore, additional independent information on the NS properties should be used. 
The detection of a gravitational wave from an NS-NS merger sheds new light on the NS equation of state \citep{Abbott17}.  The 90\% confidence limit on the tidal deformability $\Lambda<800$ of a $1.4M_\odot$ NS implies an the NS radius below $\sim$13.6 km for $M_{\rm NS}>0.5M_\odot$ \citep{Annala18}.
For the $Z=40Z_{\odot}$ model, the obtained lower limit on the NS radius of 14 km strongly contradicts this condition. 
A higher metal abundance  with $>40Z_{\odot}$ may be produced in the atmosphere during the long PRE phase of the studied burst. 
For example, the NS mass-radius constraints obtained with pure iron models, combined with the condition $R<13.6$~km, imply a 90\% confidence upper limit on the NS mass in GRS 1747$-$312 of $1.8M_{\odot}$.

\subsection{The distance  to GRS 1747$-$312/Terzan 6}
Optical measurements of the distance to Terzan 6 have been reported by several teams.  With the calculated foreground reddening to Terzan 6 $E(B-V)=2.04$, and the metallicity $\rm [Fe/H]=-0.6$ adopted from the similar globular cluster M71, the distance to Terzan 6 was measured to be $6.8\pm0.46$ kpc \citep[see][and references therein]{Fahlman95}. \citet{Barbuy97} obtained  $E(B-V)=2.24$ and $V_{\rm HB}\approx22.25$, which leads to a distance of 7 kpc to Terzan 6.   \citet{Valenti07} found $E(B-V)=2.35$ and a distance of 6.7 kpc from the color-magnitude diagram observed by the  ESO Max-Planck-Institut 2.2 m telescope.    \citet{Kuulkers03} quoted a distance  $9.5^{+3.3}_{-2.5}$ kpc to Terzan 6 from the globular cluster catalog by \citet{Harris96}, in which $E(B-V)=2.14$ was adopted. This distance estimate is larger than other measurements and the distance of 6.8 kpc given in the updated catalog by \citet{Harris10}.\footnote{\url{http://physwww.mcmaster.ca/~harris/mwgc.dat}}  
From equations ~(\ref{equ:dist}) and (\ref{equ:hb}), the foreground reddening, $E(B-V)$, contributed the largest discrepancy.

The distance to GRS 1747$-$312 can also be constrained by the direct cooling tail method and for the pure iron model we get $D=11\pm 1$ kpc (see the dotted curves in Figure~\ref{fig:MR2}), taking into account additional constraints on the radius. 
This distance is larger than that measured from optical observations, i.e., $6.7^{+2.3}_{-1.7}$ kpc. % (see Sect.~\ref{Sec:distance})
There exist a number of  possible reasons for this discrepancy. If the NS surface is partially blocked by the accretion disk, i.e., one third of NS surface is visible, the inferred distance from the direct cooling method should be 6.7 kpc instead of 11.5, while the $M-R$ contours are unchanged. But it is difficult to imagine that the disk may have such a sharp boundary in the quiescent state. The errors of  $V_{\rm HB}$, $E(B-V)$ and $\rm [Fe/H]$ are assigned rather than observed, and therefore may be underestimated. Moreover, the uncertainties in the coefficients of the empirical relations  (\ref{equ:dist}) and (\ref{equ:hb}) are not taken into account, which will lead to a larger uncertainty in the distance. Otherwise, an NS with mass smaller than $1M_\odot$ is needed.

\section{Conclusion}
\label{sec:con}

The PRE burst we studied is the most intense of all bursts from GRS 1747$-$312 \citep{Zand03b, Galloway08, Iwai14}. We find that the edge observed during the PRE phase and the cooling tail correspond to the absorption from the hydrogen-like Fe, Ni, or a mixture of Fe and Ni. The presence of the edge is confirmed by  fits to the observed flux-normalization relation with the theoretical model, which requires a high metal abundance.

We showed that the $M-R$ confidence region for even such  high metal abundance as a $40Z_{\odot}$ is inconsistent with the recent constraints on the NS radius of $<13.6$~km coming from tidal deformability during an NS-NS merger. 
However, the $M-R$ constraints obtained for pure iron are well consistent with this limit, giving an upper limit on the mass of the NS of $1.8M_{\odot}$. 
These constraints are also consistent with the gravitational redshift measurements from the absorption edge. 
The  data then limit the distance to the source to $D=11\pm 1$ kpc. 
%depend on the assumed metal abundance during the cooling tail:  
%for $Z=40Z_{\odot}$ we get $1.0-1.8M_{\odot}$, $14.0-18.0~{\rm km}$ and $10.0-13.0$ kpc, while 
%for  pure iron we obtain    $\lesssim13.5~{\rm km}$ and 
%If the NS radius constraint from GW170817, i.e., $R\lesssim13.5$ kpc, is added, the metal abundance should be higher than $40Z_{\odot}$.
Future measurements of the mass function and the type of the companion star  would help to improve the constraints on the NS mass and radius in GRS 1747$-$312.

\section*{Acknowledgments}

We appreciate the referee for the comments and suggestions for improving our work. Z.L. thank Tadayasu Dotani for the useful discussions. 
Z.L. was supported by the Swiss Government Excellence Scholarships. Z.L. thanks the International Space Science Institute and University of Bern for the hospitality. Parts of this work have been done during Z.L.'s visits to the University of Turku and the University of T\"ubingen. 
This work was supported by the National Key R\&D Program of China (No. 2017YFA0402602), the National Natural Science Foundation of China (Grant Nos. 11703021, 1673002 and U1531243), and Hunan Provincial Natural Science Foundation of China (Grant Nos. 2017JJ3310 and 2018JJ3495). 
This research has been supported in part by the grant 14.W03.31.0021 of the  Ministry of Education and Science of the Russian Federation (J.P., V.F.S.) and by the Deutsche Forschungsgemeinschaft grant WE~1312/51-1 (V.F.S.). 
%the Russian Foundation for Basic Research (grant 16-02-01145-a).
\software{Xspec (v12.8.2; \citealt{Arnaud96}).}

\bibliographystyle{apj}

\end{document}